# A new method to subdivide a spherical surface into equal-area cells

Zinovy Malkin[1,2]

[1]Pulkovo Observatory, St. Petersburg, 196140, Russia
[2]Kazan Federal University, Kazan, 420000, Russia

**Abstract.** A new method is proposed to divide a spherical surface into equal-area cells. The method is based on dividing a sphere into several latitudinal bands of near-constant span with further division of each band into equal-area cells. It is simple in construction and provides more uniform latitude step between latitudinal bands than other methods of isolatitudinal equal-area tessellation of a spherical surface.

## INTRODUCTION

Tessellating a spherical surface into equal-area cells (pixels) is a well-known problem in geodesy, astronomy, and geophysics. This operation is useful and sometimes necessary to solve different tasks, such as

- averaging data to eliminate the random error;
- resampling non-uniform data distribution over the sphere;
- optimizing computation of spherical harmonics.

Many approaches have been suggested to subdividing the earth's surface or celestial sphere into a regular way optimized for solution of different scientific and practical tasks. Analysis and comparison of these methods can be found, e.g., in Gringorten & Yepez (1992), White et al. (1992), Górski et al. (2005), Beckers & Beckers (2012), Mahdavi-Amiri et al. (2015) and papers cited therein. These methods differ from each other in several aspects:

- cell form, that can be triangle, rectangular (or equilateral spherical trapezium), non-rectangular quadrangle, pentagon, hexagon, including combination of cells of different forms;
- the cells can be or not be of the same form and dimensions;
- cell boundaries can be or not be directed or not along parallels and meridians;
- the grid can be or not be symmetric with respect to the center of the sphere.

Many applications would benefit of using isolatitudinal, rectangular, and equal-latitudinal grid. To our knowledge, no such subdivision of a spherical surface is constructed (and perhaps a rigorous solution of this problem does not exist at all). In this note, a new strategy is proposed to subdivide a sphere into (near)equal-area (near)equal-latitudinal cells.

## PROPOSED METHOD OF DIVISION OF A SPHERICAL SURFACE

In the proposed method, a spherical surface is first divided into several latitudinal (declination) bands of constant width, a division of 90 degrees. Then each band is divided into several cells of near-equal area. Cell span is selected in such a way to be about square in the equatorial bands. In other bands, the cell span is computed to be inversely proportional to the cosine of the band central latitude (declination) rounded up to the nearest divisor of 360.

The goal of further refinement of the grid is to provide equality of the cells area with the given relative accuracy required for the application for which the grid is constructed. It can be achieved by iterative adjusting the latitudinal (declination) boundaries. The cell span in the longitudinal (right ascension) direction is fixed within each latitudinal (declination) band.

Figure 1 shows three thus constructed grids consisting of 46, 130 and 406 cells corresponding to approximate latitude step 30, 18, and 10 degrees respectively. The grid parameters are given in Table 1 for the northern hemisphere. The southern hemisphere is symmetric to the northern one. The last column of the table shows the residuals of the central latitude of the latitudinal rings from the "nominal" value corresponding to the grid with the rings of equal width. Analysis of this data suggests that the residuals tend to became smaller with decreasing of the latitude step.

Table 1. Parameters of three divisions of a sphere into equal-area cells. Only the northern hemisphere is shown, the southern hemisphere is divided in the same way.

| Cells | Latitudinal (declination) bands, deg | Longitudinal (right ascension) cell span, deg | Cell area, sq. deg | Central latitude (declination), deg | | |
|---|---|---|---|---|---|---|
| | | | | Actual | Nominal | Δ |
| 46 | 60.4082 – 90 | 120 | 896.8008 | 75.20 | 75 | 0.20 |
| | 31.4490 – 60.4082 | 45 | 896.8038 | 45.93 | 45 | 0.93 |
| | 0 – 31.4490 | 30 | 896.8040 | 15.72 | 15 | 0.72 |
| 130 | 72.5246 – 90 | 120 | 317.3317 | 81.26 | 81 | 0.26 |
| | 54.6254 – 72.5246 | 40 | 317.3303 | 63.58 | 63 | 0.58 |
| | 35.7758 – 54.6254 | 24 | 317.3308 | 45.20 | 45 | 0.20 |
| | 17.9202 – 35.7758 | 20 | 317.3303 | 26.85 | 27 | –0.15 |
| | 0 – 17.9202 | 18 | 317.3302 | 8.96 | 9 | –0.04 |
| 406 | 80.1375 – 90 | 120 | 101.6085 | 85.07 | 85 | 0.07 |
| | 70.2010 – 80.1375 | 40 | 101.6082 | 75.17 | 75 | 0.17 |
| | 60.1113 – 70.2010 | 24 | 101.6083 | 65.16 | 65 | 0.16 |
| | 50.2170 – 60.1113 | 18 | 101.6077 | 55.16 | 55 | 0.16 |
| | 40.5602 – 50.2170 | 15 | 101.6084 | 45.39 | 45 | 0.39 |
| | 30.1631 – 40.5602 | 12 | 101.6084 | 35.36 | 35 | 0.36 |
| | 20.7738 – 30.1631 | 12 | 101.6086 | 25.47 | 25 | 0.47 |
| | 10.2148 – 20.7738 | 10 | 101.6086 | 15.49 | 15 | 0.49 |
| | 0 – 10.2148 | 10 | 101.6077 | 5.11 | 5 | 0.11 |

Latitudinal distribution of the cells for other isolatitudinal methods is given in Table 2. Those methods include different variants of Lambert projection (Gringorten and Yepez, 1992; Roşca, 2010; table data are taken from the first paper) and HEALPix (Górski, 2005). The number of latitudinal rings is taken to be close to variants shown in Table 1 and Figure 1. Notice that the subdivision of Gringorten and Yepez (1992) obtained using the Lambert projection includes polar cap, and HEALPix grid includes a single equatorial ring. It can be seen from Tables 1 and 2 that proposed method provides better approach to the equal-latitudinal cell distribution.

As to other methods, the hexagonal tessellation can provide equal-area and equal-latitudinal subdivision of a spherical surface. However, this grid is not hierarchical, which imposes serious limitations on its practical usefulness. Besides, it may be too complicated for some applications.

CONCLUSION

In this note, a method is presented to subdivide a spherical surface into equal-area and near-equal-latitudinal cells. A supplement advantage of this method is that the cells are rectangular with their sides directed along parallels and meridians, which makes it easy to define the pixel that contains a given point on the spherical surface. This method is simple and sufficiently accurate to meet the requirements of many practical applications. It can be used for both astronomical computations on the celestial sphere and geodesy/geophysics for computations on the Earth's surface. For example, the division into 46 cells described above was used in Malkin (2016) to resample a non-uniform distribution of radio stars over the sky to a uniformly distributed sample.

Indeed, the three grids described above are just examples of the application of the method used. These grids provide equality of the cells area with the relative accuracy better than $10^{-5}$, which looks to be sufficient for most practical applications in astronomy and geodesy. Other grids of better accuracy, as well as grids with central band and/or polar caps can be easily constructed in a similar way.

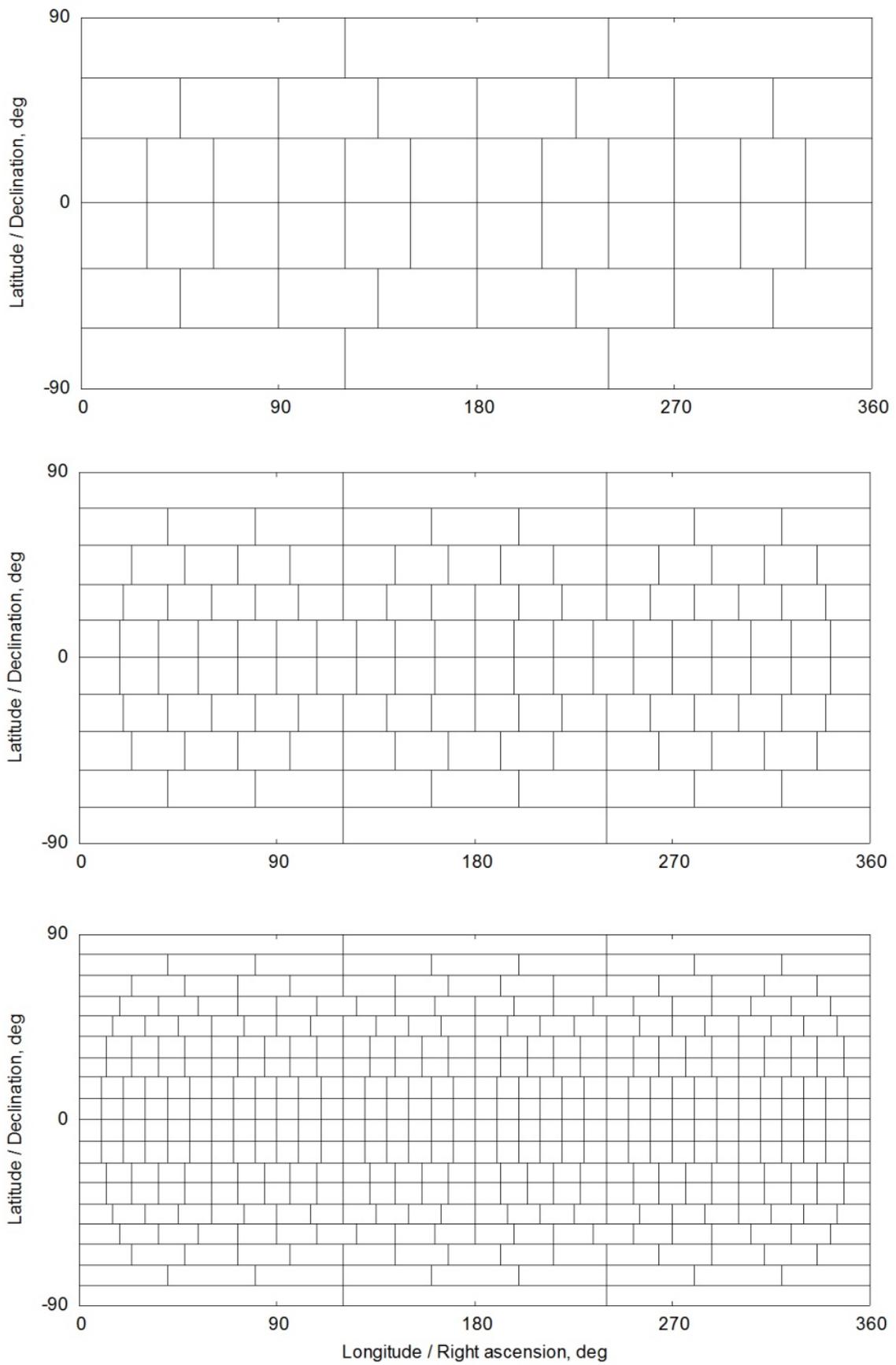

Figure 1. Dividing sphere onto 46 (top), 130 (middle), and 406 (bottom) equal-area cells.

Table 2. Central latitude of the latitudinal rings for different methods of isolatitudinal subdivision of a spherical surface. Even values correspond to the grid with the same number of rings of equal width. Only the northern hemisphere is shown, the southern hemisphere is divided in the same way.

| Lambert projection | Actual | 9.65 | 27.91 | 44.52 | 60.13 | 75.20 | | |
|---|---|---|---|---|---|---|---|---|
| | Even | 8.18 | 24.55 | 40.91 | 57.27 | 73.64 | | |
| | Δ | 1.47 | 3.36 | 3.61 | 2.86 | 1.56 | | |
| HEALPix (Nside = 2) | Actual | 66.44 | 41.81 | 19.47 | | | | |
| | Even | 77.14 | 51.43 | 25.71 | | | | |
| | Δ | –10.70 | – 9.62 | –6.24 | | | | |
| HEALPix (Nside = 4) | Actual | 78.28 | 66.44 | 54.34 | 41.81 | 30.00 | 19.47 | 9.59 |
| | Even | 84 | 72 | 60 | 48 | 36 | 24 | 12 |
| | Δ | –5.72 | –5.56 | –5.66 | –6.19 | –6.00 | –4.53 | –2.41 |